\documentclass[%
 reprint,
 amsmath,amssymb,
 aps,
 prd,
]{revtex4-1}

\usepackage{graphicx}
\usepackage{hyperref}
\usepackage[capitalise,noabbrev]{cleveref}
\usepackage{bm}
\usepackage[dvipsnames]{xcolor}
\usepackage[
    multi-part-units = single,
    range-phrase = --,
    range-units = single,
    separate-uncertainty = true
]{siunitx}

\newcommand{\Lsrc}{L_\mathrm{src}}

\begin{document}

\title{Implications of the Quantum Noise Target\\
for the Einstein Telescope Infrastructure Design}

\author{Philip Jones}
\author{Teng Zhang}
\author{Haixing Miao}
\author{Andreas Freise}

\affiliation{%
    School of Physics and Astronomy, and Institute of Gravitational Wave
    Astronomy,University of Birmingham, Edgbaston, Birmingham B15 2TT, United
    Kingdom
}

\begin{abstract}
    The design of a complex instrument such as Einstein Telescope (ET) is
    based on a target sensitivity derived from an elaborate case for
    scientific exploration. At the same time it incorporates many trade-off
    decisions to maximise the scientific value  by balancing the performance
    of the various subsystems against the cost of the installation and
    operation. In this paper we discuss the impact of a long signal recycling
    cavity (SRC) on the quantum noise performance. We show the reduction in
    sensitivity due to a long SRC for an ET high-frequency interferometer,
    provide details on possible compensations schemes and suggest a reduction
    of the SRC length. We also recall details of the trade-off between the
    length and optical losses for filter cavities, and show the strict
    requirements for an ET low-frequency interferometer. Finally, we present
    an alternative filter cavity design for an ET low-frequency interferometer
    making use of a coupled cavity, and discuss the advantages of the design
    in this context.
\end{abstract}

\date{\today}

\maketitle

\section{Introduction}

\begin{figure}[h]
    \vspace{-0em}
    \centering
    \includegraphics{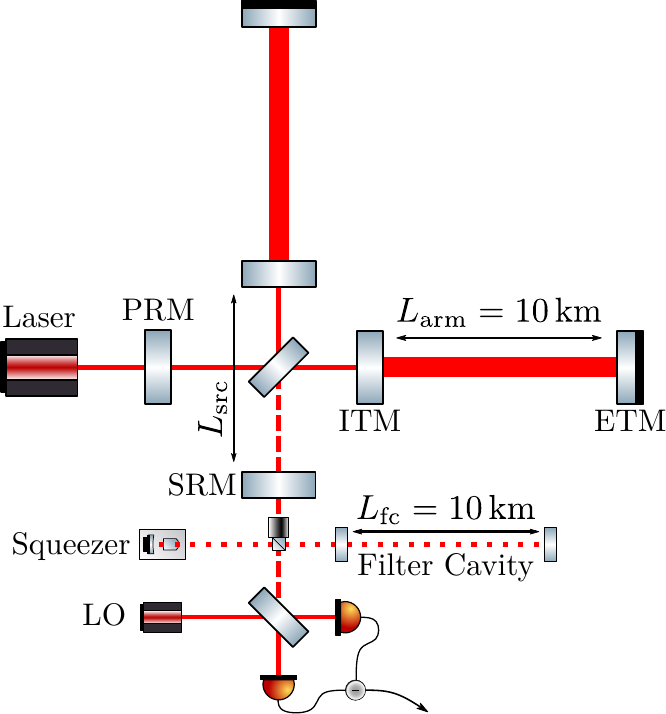}
    \centering
    \caption{Simplified interferometer design considered in this paper, with
    parameters shown for ET-HF\@. The signal recycling cavity length $\Lsrc$
    is the total distance between the signal recycling mirror (SRM) and the
    input test mass (ITM). The squeezer is a frequency-independent squeezed
    vacuum source. ET makes use of a balanced homodyne detection
    scheme, comprised of two photodiodes and a local oscillator (LO). For
    ET-LF the only addition is an extra filter cavity.}%
    \label{fig:et_hf}
\end{figure}

Current gravitational wave detectors, such as
aLIGO~\cite{TheLIGOScientific:2014jea} and Advanced
Virgo~\cite{TheVirgo:2014hva}, and plans for future detectors, such as
Einstein Telescope (ET)~\cite{et_punturo2010, Hild11, ET-D}, make use of a
dual-recycled Michelson interferometer design with arm cavities, as shown in
\cref{fig:et_hf}. There are a few key additions over a simple Michelson
interferometer, namely the power recycling mirror (PRM), the arm cavities, and
the signal recycling mirror (SRM).
The PRM acts to increase the effective input laser power; the arm cavities
increase the effective length of the arms; and the SRM reflects signal light
back into the arms, providing a way to alter the bandwidth and peak
sensitivity of the interferometer. Current detectors also make use of
frequency-independent squeezing to increase quantum-noise limited
sensitivity~\cite{PhysRevLett.110.181101, PhysRevLett.123.231107,
PhysRevLett.123.231108}. Future detectors will include frequency-dependent
squeezing to improve quantum-noise limited sensitivity, which necessitates the
addition of one or more filter cavities~\cite{PhysRevD.65.022002}. In
addition, ET features a xylophone design, where two partially overlapping
frequency ranges are investigated by different interferometer setups in the
same location~\cite{Hild:2009ns}. ET-HF operates in a tuned, broadband mode at
high frequencies of \SIrange[parse-numbers=false]{10}{10^4}{\hertz}, and ET-LF
in a detuned, narrow-band mode at low frequencies of \SIrange{1}{250}{\hertz}.

A fundamental property of interferometers is the trade-off between bandwidth
and peak sensitivity, known as the Mizuno Limit~\cite{Mizuno:1995iqa}. The
finite bandwidth of the arm cavities arises due to the gravitational wave
signal light gaining extra phase with increasing frequency (phase dispersion),
and eventually no longer resonating. This is a property of all cavities,
including the signal recycling cavity (SRC). Up until now, the length of the
SRC has been largely ignored, as it is often negligible compared to the length
of the arms ($\sim \SI{55}{\metre}$ in aLIGO, compared to \SI{4}{\kilo\metre}
arm cavities). This allowed us to treat the response of the SRC as practically
instantaneous relative to the arms, and thus the whole ITM-SRM system can be
thought of as a single compound mirror. For future detectors, especially ET,
this may no longer be the case; the initial proposed SRC length is
\SI{300}{\metre}~\cite{ET-D}. We therefore need to understand the consequences
of a non-negligible SRC length on detector design.

Another variable worth investigating is the length of the filter cavities in
ET\@. Whereas the ET design study assumed \SI{10}{\kilo\metre} long filter
cavities, in more recent discussions a reduction of this length to
\SI{1}{\kilo\metre} for ET-LF and \SI{300}{\metre} for ET-HF is being
considered. As the performance of filter cavities is determined solely by
their loss per unit length~\cite{PhysRevD.65.022002, Evans2013}, it is
necessary to understand how this reduction in length would affect the
quantum-noise limited sensitivity of ET\@.

\section{Signal Recycling Cavity Length}

There are a few motivating factors for an increased SRC length in ET compared
to that of current detectors. In the arm cavities, a relatively large beam
radius is required in order to reduce coating thermal
noise~\cite{PhysRevD.57.659}. This is especially important for ET-HF, which is
almost entirely limited by coating thermal noise around
\SIrange{40}{200}{\hertz}. At the central beamsplitter, however, a small beam
radius is desirable; it would allow smaller optics and better control of
scattered light in the central interferometer. In order to achieve such a
change in the beam sizes, a lens or telescope must be placed between the ITMs
and the beamsplitter. A short distance between the ITMs and beamsplitter, and
hence a short SRC, would require stronger focusing elements with more
stringent optics requirements to avoid introducing aberrations and noise.
Another factor leading to a long SRC is the use of cryogenic mirrors in
ET-LF\@. To achieve sufficient cooling of the ITMs, cryoshields along the
vacuum tubes are required to reduce the solid angle under which the cold ITMs
are exposed to room temperature parts of the instrument. The lengths of the
cryoshields (several tens of meters) also add to the SRC length.

\begin{figure}
    \centering
    \includegraphics{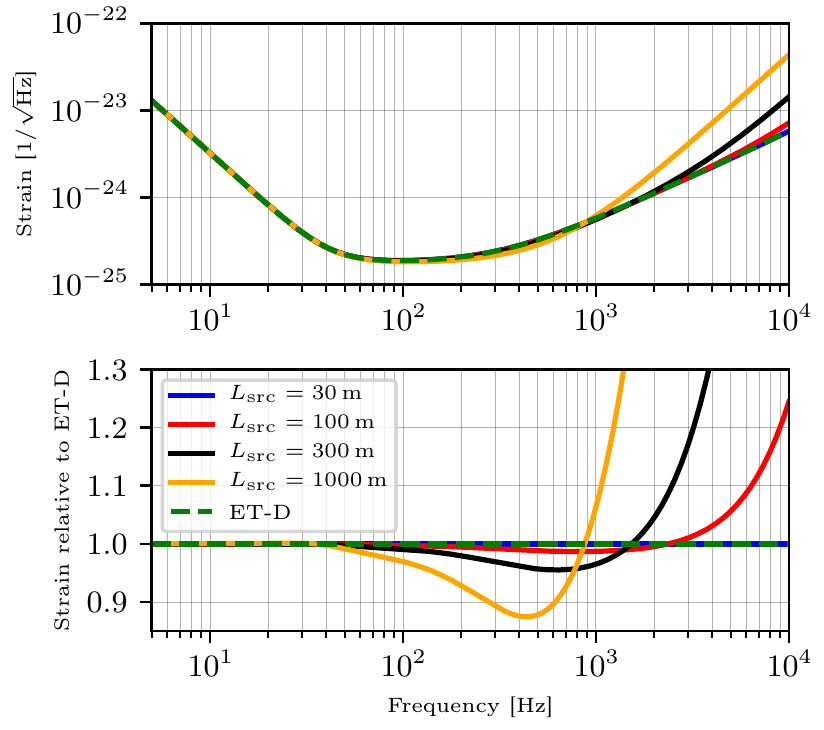}
    \caption{Effect of increasing $\Lsrc$ on ET-HF, without changing other
    interferometer parameters. For the value $\Lsrc = \SI{300}{\metre}$
    proposed in the ET Design Study~\cite{ET-D}, quantum noise deviates from
    the design curve significantly above \SI{2}{\kilo\hertz}, reaching a
    factor of 2 increase at $\sim\SI{7.5}{\kilo\hertz}$.}%
    \label{fig:et_hf_src_length}
\end{figure}

Previous models used throughout the collaboration, and in the ET Design
Study~\cite{ET-D}, assumed that the SRC length can be neglected. This is no
longer valid. We therefore need to model and understand what effects SRC
length has on the sensitivity of ET, and how to choose optimal SRC parameters
for a given length. \Cref{fig:et_hf_src_length}
shows the effects of different SRC lengths on the quantum noise performance of
the example ET-HF setup from \cref{fig:et_hf}, compared to that given in the
ET design (ET-D). All modelling throughout this paper was performed using the
frequency-domain modelling software
\textsc{Finesse}~\cite{Finesse,Freise_2004}. It should be noted that our
investigations do not yet include the effects of higher-order modes and beam
shapes. These effects should be studied further in the future, as changing the
length of the SRC impacts the detector design in other ways as well, for
example, with respect to avoiding higher-order mode resonances and parametric
instabilities~\cite{Green17}.

From \cref{fig:et_hf_src_length}, we see that increasing the SRC length
$\Lsrc$ leads mostly to a change in the gradient of the quantum noise
performance at high frequencies, with a small increase in sensitivity at
certain other frequencies. This arises due to a change in the coupled cavity
dynamics of the interferometer, and to understand how to compensate for it, we
must first understand exactly why this effect occurs.

\subsection{The SRC-Arm System as a Coupled Cavity}

For $\Lsrc \ll L_\mathrm{arm}$, where $L_\mathrm{arm}$ is the interferometer
arm length, we can treat the SRM-ITM system as a kind of compound mirror, the
only effect of which is to alter the output from the arm cavities. This is
described in {Buonanno \& Chen}~\cite{Buonanno:2002mc}, which for the tuned
SRM case gives the half-bandwidth of the SRC-arm system as
\begin{equation}
    \gamma_0 = \frac{1 + r_\mathrm{srm}}{1 - r_\mathrm{srm}}
        \gamma_\mathrm{arm},
    \label{eq:single_cavity}
\end{equation}
where $r_\mathrm{srm}$ is the amplitude reflectivity of the SRM, and
$\gamma_\mathrm{arm} = c T_\mathrm{itm} / 4 L_\mathrm{arm}$ is the
half-bandwidth of the arm cavity, with $T_\mathrm{itm}$ the power
transmissivity of the ITMs. When the SRC is comparable in length to the arm
cavities, complicated coupled cavity effects come into play. We should
therefore have a brief look at the basic properties of coupled cavities before
proceeding.

The distinguishing feature of a coupled cavity is the presence of a split
resonance. A single cavity exhibits an infinite number of equally spaced
resonances, where the frequency difference between consecutive resonances is
known as the free spectral range (FSR). A coupled cavity consists of two
cavities, each with their own FSR, and will exhibit resonance peaks whenever a
field is resonant in either of these cavities. For a field that is resonant in
both cavities (i.e.\ every common multiple of both FSRs), a split resonance
can occur, where two closely-spaced resonance peaks are observed instead of
one. In the case where the two cavities have the same length, a derivation of
the frequency difference between the two peaks is given by Th\"uring, L\"uck
and Danzmann~\cite{PhysRevE.72.066615}. If we then assume that $T_\mathrm{itm}
\ll 1$, Martynov \emph{et al.}~\cite{Martynov:2019gvu} provides formulas for
the frequency difference between the two peaks $\omega_s$, and the bandwidth
of each peak $\gamma_s$:
\begin{equation}
    \omega_s = \frac{c\sqrt{T_\mathrm{itm}}}{2\sqrt{\Lsrc L_\mathrm{arm}}},
    \qquad
    \gamma_s = \frac{c T_\mathrm{srm}}{4 \Lsrc},
    \label{eq:coupled_cavity}
\end{equation}
where $T_\mathrm{srm}$ is the power transmissivity of the SRM\@.

\begin{figure}
    \centering
    \includegraphics{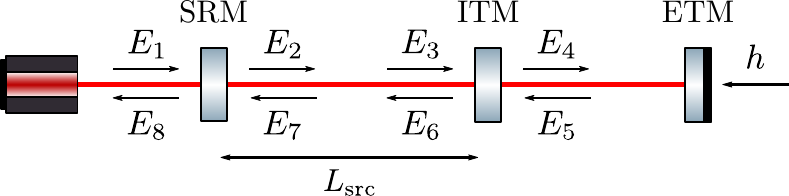}
    \caption{Simplified ET coupled cavity setup. Each of the $E_n$ are the
    full field at that point, including all frequencies. Here we are
    interested only in observing how the value of $\Lsrc$ affects the shape
    (not absolute value) of the strain-output transfer function of the
    interferometer, in a numerical simulation. In this model the only purpose of
    the laser is to provide power in the arm cavity, so it can be placed at
    either end of the setup.}%
    \label{fig:coupled_cavity}
\end{figure}

\begin{figure}
    \centering
    \includegraphics{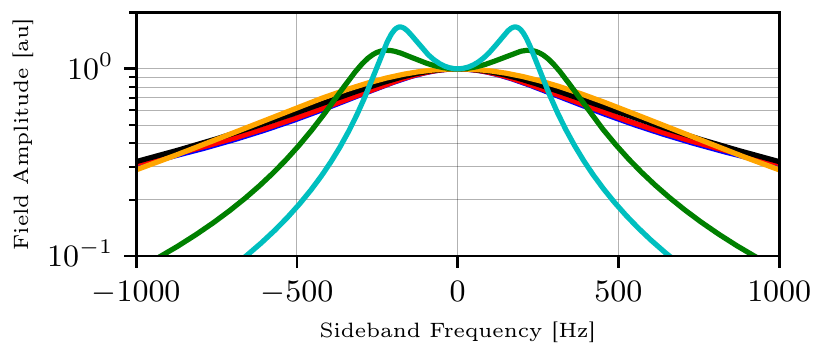}
    \begin{ruledtabular}
        \begin{tabular}{c r r r}
            \rule[-0.4\normalbaselineskip]{0pt}{1\normalbaselineskip}%
            & $\Lsrc$ [\SI{}{\metre}] & $\gamma_s$ [\SI{}{\hertz}]
            & $\omega_s$ [\SI{}{\hertz}]\\
            \hline\rule{0pt}{1\normalbaselineskip}%
            \textcolor{blue}{\rule[2pt]{2em}{2pt}}
            & 1 & \num{1.2e6} & \num{2e4}\\
            \textcolor{red}{\rule[2pt]{2em}{2pt}}
            & 100 & \num{1.2e4} & \num{2e3}\\
            \textcolor{black}{\rule[2pt]{2em}{2pt}}
            & 500 & \num{2.4e3} & \num{893}\\
            \textcolor{orange}{\rule[2pt]{2em}{2pt}}
            & 1000 & \num{1.2e3} & \num{631}\\
            \textcolor{OliveGreen}{\rule[2pt]{2em}{2pt}}
            & 5000 & \num{240} & \num{282}\\
            \textcolor{Turquoise}{\rule[2pt]{2em}{2pt}}
            & \num{10000} & \num{120} & \num{200}\\
        \end{tabular}
    \end{ruledtabular}
    \caption{Effect of SRC length on the strain-output transfer function for
    the simplified ET-HF setup shown in \cref{fig:coupled_cavity}, normalised
    to $\Lsrc = \SI{1}{\metre}$ at \SI{0}{\hertz}. As $\Lsrc$ increases,
    $\omega_s$ grows in relation to $\gamma_s$ until they are comparable in
    size and the split resonance becomes visible.}%
    \label{fig:et_coupled_cavity}
\end{figure}

For sample ET-HF parameters of $T_\mathrm{srm} = 0.1$, $T_\mathrm{itm} =
0.007$, $L_\mathrm{src} = \SI{100}{\metre}$ \& $L_\mathrm{arm} =
\SI{10}{\kilo\metre}$, we obtain $\gamma_s = \SI{12}{\kilo\hertz}$, $\omega_s
= \SI{2}{\kilo\hertz}$. As the bandwidth of the peaks is much greater than the
separation, they are indistinguishable, and the whole system has a single
resonance peak per FSR\@. However, as $\Lsrc$ increases, $\gamma_s$ decreases
faster than $\omega_s$, and the two peaks become more resolved. For the
simplified setup in \cref{fig:coupled_cavity},
the effects of increasing $\Lsrc$ are shown in \Cref{fig:et_coupled_cavity}.

It should be noted that \cref{eq:single_cavity,eq:coupled_cavity} are only
correct in their relevant extremes. In reality, interferometers operate
somewhere between the two, where although calculating the response of the
setup in \cref{fig:coupled_cavity} is simple, there is no analytical solution
for the bandwidth of the one or two peaks present. However,
\cref{eq:single_cavity,eq:coupled_cavity} are useful as a starting point for
investigating the behaviour of a long SRC with numerical simulations.

\begin{figure}
    \centering
    \includegraphics{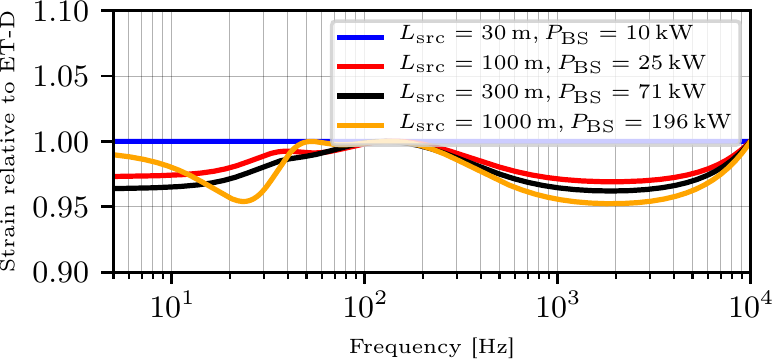}
    \vspace{.5em}
    \begin{ruledtabular}
        \begin{tabular}{c r r r r r}
            \rule[-0.4\normalbaselineskip]{0pt}{1\normalbaselineskip}%
            & $\Lsrc$ [\SI{}{\metre}] & $P_\mathrm{BS}$ [\SI{}{\kilo\watt}]
            & $P_\mathrm{in}$ [\SI{}{\watt}] & $T_\mathrm{itm}$
            [\SI{}{\percent}]
            & $T_\mathrm{srm}$ [\SI{}{\percent}]\\
            \hline\rule{0pt}{1\normalbaselineskip}%
            \textcolor{blue}{\rule[2pt]{2em}{2pt}}
            & 30 & \num{10} & 504 & \num{0.70} & \num{10.0} \\
            \textcolor{red}{\rule[2pt]{2em}{2pt}}
            & 100 & \num{25} & 614 & \num{1.68} & \num{21.9}\\
            \textcolor{black}{\rule[2pt]{2em}{2pt}}
            & 300 & \num{71} & 1141 & \num{4.62} & \num{50.2}\\
            \textcolor{orange}{\rule[2pt]{2em}{2pt}}
            & 1000 & \num{196} & 2661 & \num{12.25} & \num{86.8}
        \end{tabular}
    \end{ruledtabular}
    \caption{Options for correcting for an increased SRC length in ET-HF\@.
    Parameters were found by numerically optimising to give the minimum
    beamsplitter power possible, without decreasing sensitivity relative to
    ET-D. Compared to the uncorrected curves in \cref{fig:et_hf_src_length},
    we no longer see a decrease in sensitivity at high frequency with
    increasing $\Lsrc$, at the cost of increased power on the central
    beamsplitter---the minimum power required to keep below the original curve
    scales slightly less than linearly with SRC length. It is noteworthy that
    SRM transmissivity places a physical limit on how much $\gamma_s$ can be
    increased for a given value of $\Lsrc$. For $\Lsrc > \SI{1000}{\metre}$,
    this would start to constrain the design more strictly. It is also notable
    that the scaling relations from \cref{eq:coupled_cavity} do not hold over
    the whole range of lengths shown; for $\Lsrc = \SI{1000}{\metre}$, the
    required mirror transmissivities are much lower than one would expect.}%
    \label{fig:et_src_length_compensation}
\end{figure}

\begin{figure}
    \centering
    \includegraphics{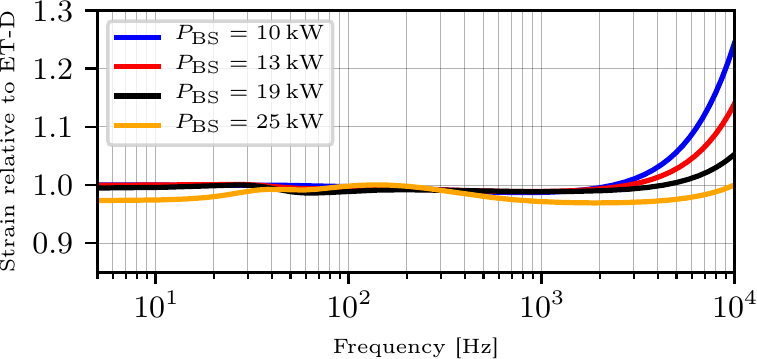}
    \caption{Scaling of quantum-noise limited sensitivity with power incident
    on the central beamsplitter for a fixed SRC length of \SI{100}{\metre} in
    ET-HF\@. For future high-power detector designs with significant SRC
    lengths, we can consider the trade-off between slightly reduced
    sensitivity due to quantum noise at high frequency, and increased noise
    due to thermal effects when compensating for SRC length.}%
    \label{fig:et_src_pbs_scaling}
\end{figure}

It should now be clear why we see a decrease in sensitivity at high frequency
for longer SRC lengths in \cref{fig:et_hf_src_length}, as increasing $\Lsrc$
reduces the bandwidth of the coupled cavity resonance and decreases the
magnitude of the frequency-splitting. To combat this, we can restore
$\omega_s$ \& $\gamma_s$ to their original values, or as close as possible.
For a change in SRC length from $\Lsrc$ to $\Lsrc'$, we should therefore
increase $T_\mathrm{srm}$ \& $T_\mathrm{itm}$ by the same ratio $\Lsrc' /
\Lsrc$. By increasing $T_\mathrm{itm}$, however, we change both the finesse of
the arm cavities, and the gain of the power recycling cavity (PRC). This
reduces the circulating arm power, and also redistributes power in the
interferometer from the arm cavities to the PRC\@. If we then increase input
power to restore the arm cavity circulating power, we can recover the original
quantum noise sensitivity curve with a larger $\Lsrc$, at the cost of
increased power incident on the central beamsplitter and transmitted through
the ITMs. This is undesirable as absorbed laser power causes thermal
distortion of the optics, creating a thermal lens which can lead to mode
mismatches and losses. Compensation for different values of $\Lsrc$, along
with power incident on the beamsplitter, is shown in
\cref{fig:et_src_length_compensation}.
A good compromise for ET-HF, including the beam expansion telescope, can be
achieved with $\Lsrc$ of around \SI{100}{\metre}.
\Cref{fig:et_src_pbs_scaling} shows how the quantum noise at high frequencies
scales with power incident on the central beamsplitter for this length.

\begin{figure}
    \centering
    \includegraphics{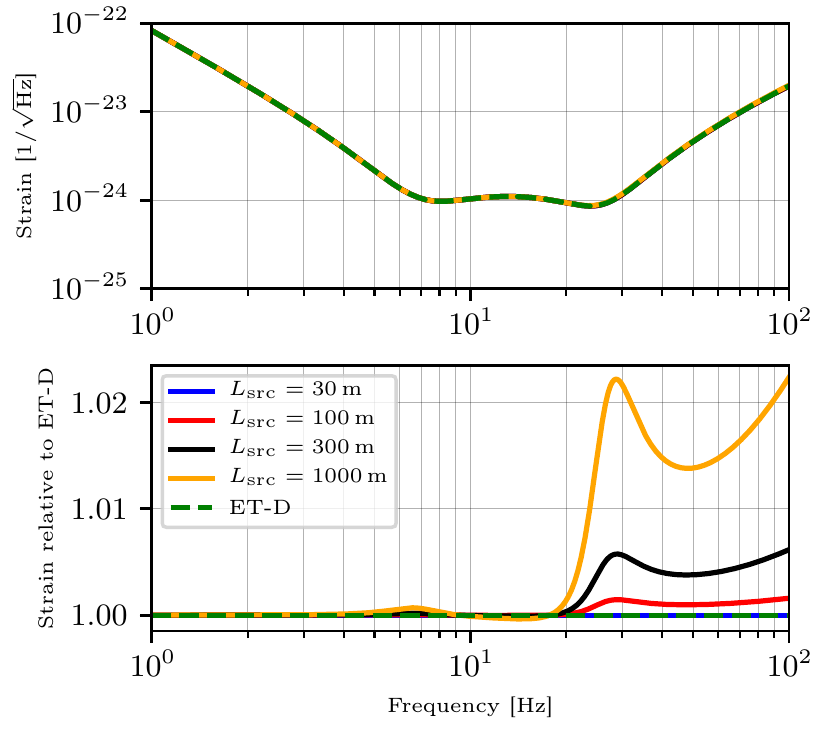}
    \caption{Effect of increasing $\Lsrc$ on ET-LF, without changing any other
    interferometer parameters. We see that even for $\Lsrc =
    \SI{1}{\kilo\metre}$, the reduction in sensitivity is only
    \SI{2}{\percent} at \SI{30}{\hertz}, and for a realistic length of
    \SI{100}{\metre}, the effect is on the order of \SI{0.1}{\percent}.}%
    \label{fig:et_lf_src_length}
\end{figure}

So far we have only discussed the effect of increasing $\Lsrc$ on ET-HF\@.
This is because, for any practical value of $\Lsrc \lesssim
\SI{1}{\kilo\metre}$, the effect on the frequency range of interest for ET-LF
(up to $\approx \SI{30}{\hertz}$) is negligible; this is shown explicitly in
\cref{fig:et_lf_src_length}. \Cref{fig:et_coupled_cavity} provides the
explanation for this behaviour. For a \SI{1}{\kilo\metre} SRC, we have
$\gamma_s = \SI{1.2}{\kilo\hertz},\ \omega_s = \SI{631}{\hertz}$---the split
resonances are still too wide to be individually resolved, and the splitting
frequency is much greater than the top end of the frequency range of interest.

\section{Optimised Filter Cavities for ET}

In order to produce frequency-dependent squeezing to improve the quantum-noise
limited sensitivity of ET, frequency-independent squeezed light is reflected
from one or more filter cavities. This induces a frequency-dependent phase
shift in the reflected light. The ET Design Study~\cite{ET-D} considered
\SI{10}{\kilo\metre} long filter cavities. Shorter filter cavities are under
consideration as a cost saving change to the design, as the vacuum and tunnel
infrastructure are one of the main costs of the future observatory.
Significantly shortened filter cavities would allow a simplification of the
infrastructure design---example lengths being considered are a reduction from
\SI{10}{\kilo\metre} down to \SI{1}{\kilo\metre} for ET-LF, and
\SI{300}{\metre} for ET-HF\@. The performance of filter cavities is determined
by their loss per unit length~\cite{PhysRevD.65.022002}, and this reduction in
length will lead to a corresponding increase in squeezing loss in the filter
cavities. In practice, the optical loss will be determined by the detailed
properties of the optical surface and the beam radius~\cite{Isogai:2013wfa},
the minimum value of which is dependent on cavity length. A simple
extrapolation from other experiments would suggest the following optical
losses are achievable with current technology and techniques for the different
filter cavity lengths~\cite{Evans2013}: \SI{30}{ppm} @ \SI{300}{\metre},
\SI{40}{ppm} @ \SI{1}{\kilo\metre} and \SI{75}{ppm} @ \SI{10}{\kilo\metre}.
Throughout this section, we do not attempt to predict detailed optical losses,
but provide quantum-noise limited sensitivity curves for a range of possible
round-trip filter cavity power losses.

\subsection{ET-HF}

For ET-HF in its tuned, broadband configuration, only one filter cavity is
required. In this case, an analytical solution for the optimal filter cavity
detuning and bandwidth is given
in~\cite[Equations (31, 33, 49, 50 \& 53)]{Kwee:2014vba} as
\begin{align}
    \Delta\omega_\mathrm{fc} &=
        \sqrt{1 - \epsilon}\gamma_\mathrm{fc},\label{eq:omega_fc}\\
    \gamma_\mathrm{fc} &=
        \sqrt{\frac{2}{(2 - \epsilon)\sqrt{1 - \epsilon}}}
        \frac{\Omega_\mathrm{SQL}}{2},\label{eq:gamma_fc}\\
    \text{where } \epsilon &=
        \frac{4}{%
            2 + \sqrt{%
                2 + 2\sqrt{%
                    1 + {\left(
                        \frac{%
                            2\Omega_\mathrm{SQL}
                        }{%
                            f_\mathrm{FSR}\Lambda_\mathrm{rt}^2
                        }
                    \right)}^4
                }
            }
        }\\
    \text{and } \Omega_\mathrm{SQL} &\simeq
        \frac{t_\mathrm{srm}}{1 + r_\mathrm{srm}}
        \frac{8}{c}
        \sqrt{\frac{P_\mathrm{arm}\omega_0}{m T_\mathrm{itm}}}.
\end{align}
Here, $f_\mathrm{FSR}$ is the free spectral range of the filter cavity,
$\Lambda_\mathrm{rt}^2$ is the round-trip power loss in the filter cavity,
$t_\mathrm{srm}$ \& $r_\mathrm{srm}$ are the amplitude transmissivity and
reflectivity of the signal recycling mirror, $P_\mathrm{arm}$ is the
circulating power in the arm cavities, $\omega_0$ is the carrier frequency,
$m$ is the mass of the test masses, and $T_\mathrm{itm}$ is the power
transmissivity of the input test mass mirror.

\begin{figure}
    \centering
    \includegraphics{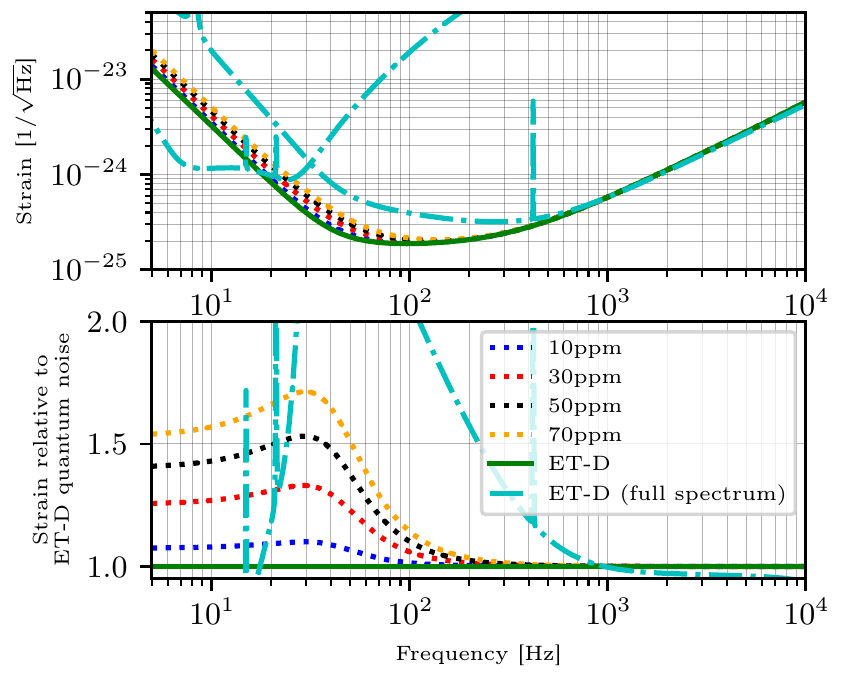}
    \caption{The effect of filter cavity loss on ET-HF, with $L_\mathrm{fc} =
    \SI{300}{\metre}$. For a round-trip loss of \SI{70}{ppm}, we see an
    increase in quantum noise of $\simeq \SI{75}{\percent}$ at
    \SI{30}{\hertz}, relative to that of ET-D. However, ET-D is limited by
    thermal noise around this frequency, so the overall sensitivity is not
    affected. As filter cavity performance is determined by loss per unit
    length, the lower loss curves can be used to infer curves for other
    lengths e.g.\ a \SI{700}{\metre} filter cavity with \SI{70}{ppm}
    round-trip loss would give the \SI{30}{ppm} curve shown.}%
    \label{fig:et_hf_loss_300m}
\end{figure}
\Cref{fig:et_hf_loss_300m} shows the effect of loss on ET-HF, with filter
cavity length $L_\mathrm{fc} = \SI{300}{\metre}$, and
\cref{tab:et_hf_fc_params}
\begin{table}
    \centering
    \begin{ruledtabular}
        \begin{tabular}{c c c}
            \rule[-0.4\normalbaselineskip]{0pt}{1\normalbaselineskip}%
            $L_\mathrm{fc}$ [\SI{}{\metre}]
            & Tuning [\SI{}{\hertz}]
            & Half-bandwidth [\SI{}{\hertz}] \\
            \hline\rule{0pt}{1\normalbaselineskip}%
            300 & -29.9520 & 5.2305 \\
            %500 & -29.9932 & 5.0467 \\
        \end{tabular}
    \end{ruledtabular}
    \caption{Optimal filter cavity parameters for ET-HF, with \SI{70}{ppm}
    round-trip filter cavity loss.}%
    \label{tab:et_hf_fc_params}
\end{table}
gives the optimal filter cavity parameters according to
\cref{eq:omega_fc,eq:gamma_fc}. The effect of the increased losses is a
reduction in the quantum-noise limited sensitivity at low frequencies,
especially around \SI{30}{\hertz}. However, in this frequency band the current
ET-HF design is entirely limited by thermal noise, so the overall sensitivity
is not affected. We expect that upgrades to the interferometers in the
long-term infrastructure of ET will reduce this thermal noise. Space for a
filter cavity of modest length should therefore already be allocated in the
initial infrastructure.

\subsection{ET-LF}%
\label{sec:et_lf_sqz}

ET-LF operates with a detuned SRM, and thus requires two filter cavities to
achieve optimal squeezing. Unlike in the single filter cavity case, no
analytical solution exists for multiple lossy filter cavities. A good
approximation is provided by~\cite[Appendix A]{Purdue:2002md}, however this
assumes lossless filter cavities. Thus, we start with this approximation, and
then optimise numerically.
\begin{figure}
    \centering
    \includegraphics{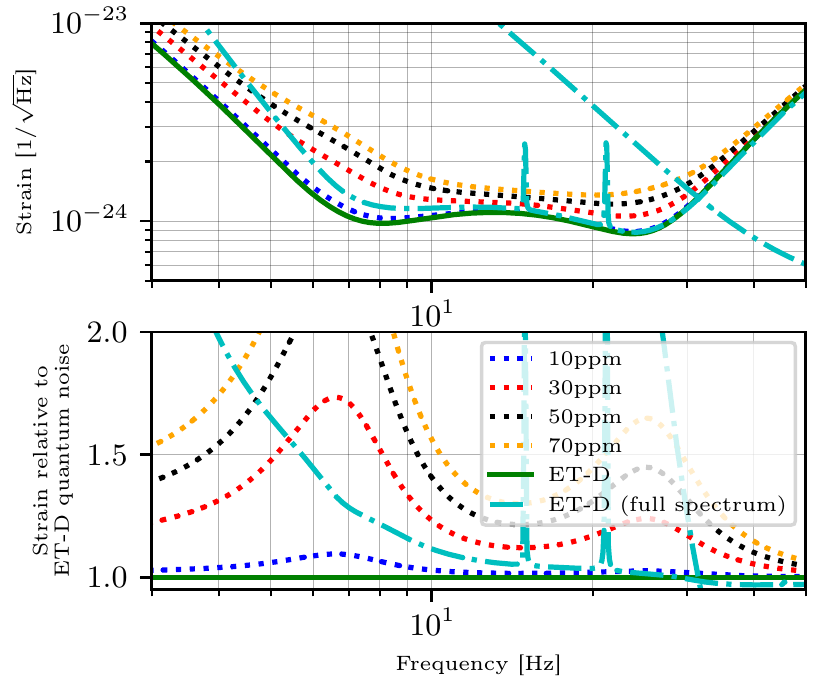}
    \caption{The effect of filter cavity loss on ET-LF, with $L_\mathrm{fc} =
    \SI{1}{\kilo\metre}$. For a round-trip loss of
    \SI{70}{ppm}, we see an increase in quantum noise of  up to $
    \SI{160}{\percent}$ at \SI{7}{\hertz}, relative to that of ET-D
    (\SI{75}{ppm} loss, $L_\mathrm{fc} = \SI{10}{\kilo\metre}$), and a minimum
    of \SI{30}{\percent} over the full spectrum below \SI{30}{\hertz}.}%
    \label{fig:et_lf_loss_1000m}
\end{figure}%
\Cref{fig:et_lf_loss_1000m} shows the effects of losses on
ET-LF, with $L_\mathrm{fc} = \SI{1}{\kilo\metre}$, and
\cref{tab:et_lf_fc_params}
\begin{table}
    \centering
    \begin{ruledtabular}
        \begin{tabular}{c c c}
            \rule[-0.4\normalbaselineskip]{0pt}{1\normalbaselineskip}%
            Filter & Tuning [\SI{}{\hertz}] & Half-bandwidth [\SI{}{\hertz}] \\
            \hline\rule{0pt}{1\normalbaselineskip}%
            FC$_1$ & 25.3574 & 5.6830 \\
            FC$_2$ & -6.6366 & 1.4468 \\
        \end{tabular}
    \end{ruledtabular}
    \caption{Optimal filter cavity parameters for ET-LF, with \SI{70}{ppm}
    round-trip filter cavity loss.}%
    \label{tab:et_lf_fc_params}
\end{table}
gives optimal filter cavity parameters. We see that a reduction in length
leads to a much more stringent requirement for the optical loss to avoid
spoiling the sensitivity at low frequencies around \SI{7}{\hertz}. The Virgo
detector has already demonstrated round-trip losses of \SI{55(10)}{ppm} in
km-scale cavities~\cite{degallaix_2019}. These optical losses are dominated by
deficiencies in the mirror surface quality, and research is ongoing to
identify and mitigate the loss due to light scattering by surface defects.
Using the same technology it should currently be possible to realise a
\SI{1}{\kilo\metre} long cavity with round-trip losses of less than
\SI{40}{ppm}. It is reasonable to believe that in the future we can improve
the mirror surface quality further, to achieve a round trip loss of
\SI{20}{ppm}, with careful use of state-of-the-art technologies and care
regarding polishing, coating, handling and installation.

\begin{figure}
    \centering
    \includegraphics{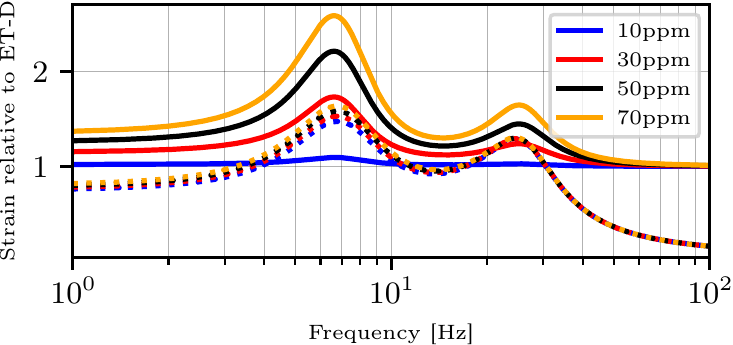}
    \caption{Comparison of a detuned ET-LF with $2 \times \SI{1}{\kilo\metre}$
    filter cavities (solid curves) and a tuned ET-LF with
    $1\times\SI{10}{\kilo\metre}$ filter cavity (dotted curves). For a
    round-trip power loss $>\SI{30}{ppm}$, the tuned system with one filter
    cavity performs better across the entire frequency range, especially at
    high frequencies.}%
    \label{fig:et_lf_tuned_vs_detuned}
    \vspace{-.7em}
\end{figure}

The main motivation for reducing the filter cavity lengths in ET-LF is the
cost of the infrastructure. There is a disincentive to use the main
\SI{10}{\kilo\metre} tunnels for the arm cavities as well as two filter
cavities, due to the scaling of excavation cost with tunnel diameter. Given
the lack of a low-frequency dip for a \SI{1}{\kilo\metre} filter cavity with loss $\gtrsim
\SI{20}{ppm}$, it is interesting to compare the previous results to the
performance of a tuned SRM for ET-LF with only one filter cavity. A comparison
of this tuned ET-LF with a single \SI{10}{\kilo\metre} long filter cavity, and
a detuned ET-LF with two \SI{1}{\kilo\metre} long filter cavities, is shown in
\cref{fig:et_lf_tuned_vs_detuned}. For a round-trip power loss
$>\SI{30}{ppm}$, the tuned system with one filter cavity performs better than
the detuned system with two short filter cavities.

The design for the ET-LF filter cavity scheme is more complex than for ET-HF,
and has to include a careful trade-off between excavation cost, expected
optical losses and practical constraints for arranging the vacuum system.

\subsection{Coupled Filter Cavities}

The purpose of using filter cavities in the squeezing path is to replicate the
quadrature rotation of the interferometer as seen by the signal light. For a
detuned signal-recycled Michelson such as ET-LF, two separate rotations are
required. Current plans for ET-LF achieve the desired rotation with two
separate filter cavities in series, with each filter cavity producing a single
rotation around its resonance. As a coupled cavity exhibits two separate
resonances, these two independent filter cavities could potentially be
replaced with a coupled filter cavity. To investigate this possibility, a
model of a coupled filter cavity was numerically fit to give the same
squeezing angle rotation as the two filter cavities, and then further
optimised to maximise the quantum-noise limited sensitivity from
\SIrange{5}{30}{\hertz}.

\begin{figure}
    \centering
    \includegraphics{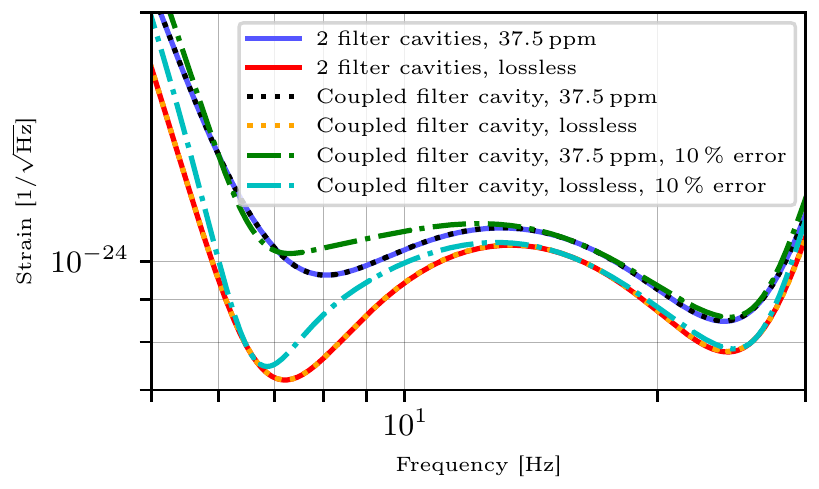}
    \caption{Comparison of using two separate \SI{10}{\kilo\metre} filter
    cavities vs.\ a coupled filter cavity totalling \SI{20}{\kilo\metre} in
    ET-LF, both with lossless mirrors and \SI{37.5}{ppm} loss per surface. The
    parameter with the lowest tolerance is the transmissivity of the middle
    mirror; the dash-dotted lines show the coupled cavity with a
    \SI{10}{\percent} increase in the middle mirror transmissivity from the
    optimal value, and the other parameters adjusted to compensate. A coupled
    filter cavity can be used to replicate the effects of two independent
    filter cavities in ET-LF\@. Perhaps surprisingly, the performance in the
    presence of losses is identical for the same loss per surface.}%
    \label{fig:et_lf_coupled_fc}
\end{figure}

\Cref{fig:et_lf_coupled_fc}
compares the quantum-noise limited sensitivity of ET-LF for a
\SI{20}{\kilo\metre} coupled filter cavity vs\
\SI[parse-numbers=false]{2\times10}{\kilo\metre} filter cavities, both with
and without losses. Optimal parameters are given in
\cref{tab:et_lf_coupled_fc_params}.
\begin{table*}
    \centering
    \begin{ruledtabular}
        \begin{tabular}{c r r r r}
            \rule[-0.4\normalbaselineskip]{0pt}{1\normalbaselineskip}%
            Type
            & \multicolumn{2}{c}{Transmissivity}
            & \multicolumn{2}{c}{Tuning [\SI{}{degrees}]} \\
            \hline\rule{0pt}{1\normalbaselineskip}%
            %[-2.66665000e-01, -3.94714000e-01,  1.94085828e-02,  2.57993124e-05]
            %[2.25068645e-01 2.25756139e-01 6.03001983e-03 3.11499412e-05]
            Two cavities
            & FC\textsubscript{1,in} = \SI{4.617e-3}{}
            & FC\textsubscript{2,in} = \SI{1.210e-3}{}
            & FC\textsubscript{1,end} = \SI{3.049e-1}{}
            & FC\textsubscript{2, end} = \SI{-7.971e-2}{}
            \\
            Coupled cavity
            & FC\textsubscript{in} = \SI{5.856e-3}{}
            & FC\textsubscript{mid} = \SI{3.099e-5}{}
            & FC\textsubscript{mid} = \SI{2.276e-1}{}
            & FC\textsubscript{end} = \SI{2.256e-1}{}
            \\
        \end{tabular}
    \end{ruledtabular}
    \caption{Optimal filter cavity parameters for ET-LF, for both two filter
    cavities and a single coupled filter cavity. Each individual cavity is
    \SI{10}{\kilo\metre} long, with \SI{37.5}{ppm} loss per optic. Note the
    low transmissivity required for the middle mirror in the coupled filter
    cavity. As cavity length decreases, so too does the required value of
    middle mirror transmissivity.}%
    \label{tab:et_lf_coupled_fc_params}
\end{table*}
There are a few noteworthy points here. Firstly, from
\cref{fig:et_lf_coupled_fc}, we can see that the fit was performed
successfully, and as such a coupled filter cavity could in theory be used in
place of two independent cavities in a detuned, dual-recycled Michelson such
as ET-LF\@. Additionally, we see that the scaling of performance with mirror
losses is identical in the coupled and independent cases. We also see that the
performance of the filter cavity at low frequencies
($\sim$\SIrange{7}{12}{\hertz}) is fairly sensitive to the middle mirror
transmissivity.

\begin{figure}
    \centering
    \includegraphics{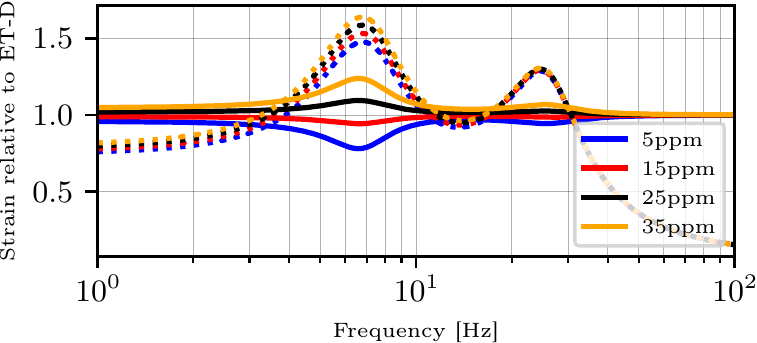}
    \caption{Comparison of a detuned ET-LF with a \SI{10}{\kilo\metre} total
    length coupled filter cavity (solid curves) and a tuned ET-LF with a
    single \SI{10}{\kilo\metre} filter cavity (dotted curves). For all values
    of per-surface power loss plotted, the coupled filter cavity leads to
    better average sensitivity from $\sim$\SIrange{4}{30}{\hertz}, and a loss
    of \SI{20}{ppm} brings the sensitivity in line with that of ET-D.}%
    \label{fig:et_lf_tuned_vs_coupled_detuned}
\end{figure}

There are a few motivating factors that make the coupled filter cavity design
worth further study. When two individual cavities are used, some extra optics
such as Faraday isolators must be introduced to direct the beam from one
cavity to the next; this is not needed in a coupled filter cavity. Without
these extra Faraday isolators the overall optical loss in the input squeezing
path can be reduced~\cite{PhysRevLett.123.231107}, to increase the effective
squeezing level achievable. Additionally, the same considerations for using a
tuned vs.\ detuned Michelson for ET-LF from \cref{sec:et_lf_sqz} apply here:
in the case of the coupled cavity scheme, the total filter cavity length is
arranged sequentially in one long vacuum system, whereas the scheme with two
filter cavities requires a shorter but wider space for two parallel vacuum
systems. It should be noted that a coupled filter cavity of total length
\SI{10}{\kilo\metre} could provide much better sensitivity than a tuned
detector in each of ET-LF's `dips'. This is shown explicitly in
\cref{fig:et_lf_tuned_vs_coupled_detuned}.

In summary, a coupled filter cavity could be used in place of two independent
filter cavities. There are two main advantages to this substitution: the lack
of a need for an extra Faraday isolator results in lower losses in the
squeezing path, and in the case of ET the form factor of the vacuum system
could be advantageous. These advantages provide motivation for further study.

\section{Summary}

Practical considerations may motivate the introduction of longer signal
recycling cavities in future detectors, such as ET\@. If the length of the SRC
is not accounted for, it can lead to an overestimation of the sensitivity at
high frequencies, as shown in \cref{fig:et_hf_src_length}. This can be avoided
by considering the SRC-arm system as a coupled cavity, the response of which
is described loosely by a split resonance, with separation frequency
$\omega_s$ and half-bandwidth $\gamma_s$, as given in
\cref{eq:coupled_cavity}. We have shown that the change in the response of a
detector due to a longer SRC can be counteracted by increasing both
$T_\mathrm{itm}$ \& $T_\mathrm{srm}$. To maintain arm cavity power and thus
sensitivity, we must also increase input power, either directly or by
increasing the PRC gain. Increasing the length of the SRC, while maintaining
sensitivity at high frequencies, therefore leads to an increase in the power
incident on the central beamsplitter, as shown in
\cref{fig:et_src_length_compensation}.

In the specific case of ET-HF, we suggest an SRC length of \SI{100}{\metre},
which would result in a reduction of the quantum-noise limited sensitivity of
only \SI{25}{\percent} at \SI{10}{\kilo\hertz} compared to an SRC of negligible
length. We further show how this loss of sensitivity can be compensated for, at
the cost of increased laser power at the beamsplitter. The scaling of this
sensitivity with beamsplitter power can be seen in
\cref{fig:et_src_pbs_scaling}. For ET-LF, the frequencies at which the
decrease in sensitivity occurs are too high to be of consequence.
\Cref{fig:et_lf_src_length} shows that for a \SI{100}{\metre} SRC, the
reduction in sensitivity at \SI{30}{\hertz} is on the order of
\SI{0.1}{\percent}.

Constraints such as the size and cost of the underground infrastructure
provide motivation for reducing the length of the filter cavities used for
frequency-dependent squeezing in ET\@. This has the effect of increasing
filter cavity loss per unit length, leading to a reduction in performance. For
ET-HF, a much shorter filter cavity of, for example, \SI{300}{\metre} has
limited consequences, as thermal noise of the main interferometer remains the
limiting factor in the frequency range affected by filter cavity losses. In
ET-LF, the consequences of reducing the filter cavity lengths from
\SI{10}{\kilo\metre} to \SI{1}{\kilo\metre} would be more severe, giving up to
a factor of 2 reduction in quantum-noise limited sensitivity at \SI{7}{\hertz}
for a currently achievable round-trip optical power loss of \SI{40}{ppm}.
However, with expected improvements in optical losses such as an increase in
mirror surface quality, a significant reduction of the filter cavity length
would be possible. In addition, the use of a coupled filter cavity in place of
two independent filter cavities for ET-LF was investigated, and found to
perform identically for the same length and per-surface loss. Combined with
the fact that coupled cavities would use one less Faraday isolator in the
injection path, which further reduces the optical losses, coupled filter
cavities should be studied further.

\begin{acknowledgments}
    We are grateful to J\'er\^ome Degallaix for providing information on
    optical losses and Jan Harms and Harald L\"uck for helpful comments. This
    work has been supported by the Science and Technology Facilities Council
    (STFC), A. Freise acknowledges support from a Royal Society Wolfson
    Fellowship which is jointly funded by the Royal Society and the Wolfson
    Foundation. This paper has the LIGO Document Number LIGO-P2000066.
\end{acknowledgments}


\begin{thebibliography}{23}%
\makeatletter
\providecommand \@ifxundefined [1]{%
 \@ifx{#1\undefined}
}%
\providecommand \@ifnum [1]{%
 \ifnum #1\expandafter \@firstoftwo
 \else \expandafter \@secondoftwo
 \fi
}%
\providecommand \@ifx [1]{%
 \ifx #1\expandafter \@firstoftwo
 \else \expandafter \@secondoftwo
 \fi
}%
\providecommand \natexlab [1]{#1}%
\providecommand \enquote  [1]{``#1''}%
\providecommand \bibnamefont  [1]{#1}%
\providecommand \bibfnamefont [1]{#1}%
\providecommand \citenamefont [1]{#1}%
\providecommand \href@noop [0]{\@secondoftwo}%
\providecommand \href [0]{\begingroup \@sanitize@url \@href}%
\providecommand \@href[1]{\@@startlink{#1}\@@href}%
\providecommand \@@href[1]{\endgroup#1\@@endlink}%
\providecommand \@sanitize@url [0]{\catcode `\\12\catcode `\$12\catcode
  `\&12\catcode `\#12\catcode `\^12\catcode `\_12\catcode `\%12\relax}%
\providecommand \@@startlink[1]{}%
\providecommand \@@endlink[0]{}%
\providecommand \url  [0]{\begingroup\@sanitize@url \@url }%
\providecommand \@url [1]{\endgroup\@href {#1}{\urlprefix }}%
\providecommand \urlprefix  [0]{URL }%
\providecommand \Eprint [0]{\href }%
\providecommand \doibase [0]{https://doi.org/}%
\providecommand \selectlanguage [0]{\@gobble}%
\providecommand \bibinfo  [0]{\@secondoftwo}%
\providecommand \bibfield  [0]{\@secondoftwo}%
\providecommand \translation [1]{[#1]}%
\providecommand \BibitemOpen [0]{}%
\providecommand \bibitemStop [0]{}%
\providecommand \bibitemNoStop [0]{.\EOS\space}%
\providecommand \EOS [0]{\spacefactor3000\relax}%
\providecommand \BibitemShut  [1]{\csname bibitem#1\endcsname}%
\let\auto@bib@innerbib\@empty
%</preamble>
\bibitem [{\citenamefont {Aasi}\ \emph {et~al.}(2015)\citenamefont {Aasi} \emph
  {et~al.}}]{TheLIGOScientific:2014jea}%
  \BibitemOpen
  \bibfield  {author} {\bibinfo {author} {\bibfnamefont {J.}~\bibnamefont
  {Aasi}} \emph {et~al.} (\bibinfo {collaboration} {LIGO Scientific
  Collaboration}),\ }\href {https://doi.org/10.1088/0264-9381/32/7/074001}
  {\bibfield  {journal} {\bibinfo  {journal} {Class. Quant. Grav.}\ }\textbf
  {\bibinfo {volume} {32}},\ \bibinfo {pages} {074001} (\bibinfo {year}
  {2015})},\ \Eprint {https://arxiv.org/abs/1411.4547} {arXiv:1411.4547
  [gr-qc]} \BibitemShut {NoStop}%
%%CITATION = ARXIV:1411.4547;%%
\bibitem [{\citenamefont {Acernese}\ \emph {et~al.}(2015)\citenamefont
  {Acernese} \emph {et~al.}}]{TheVirgo:2014hva}%
  \BibitemOpen
  \bibfield  {author} {\bibinfo {author} {\bibfnamefont {F.}~\bibnamefont
  {Acernese}} \emph {et~al.} (\bibinfo {collaboration} {Virgo Collaboration}),\
  }\href {https://doi.org/10.1088/0264-9381/32/2/024001} {\bibfield  {journal}
  {\bibinfo  {journal} {Class. Quant. Grav.}\ }\textbf {\bibinfo {volume}
  {32}},\ \bibinfo {pages} {024001} (\bibinfo {year} {2015})},\ \Eprint
  {https://arxiv.org/abs/1408.3978} {arXiv:1408.3978 [gr-qc]} \BibitemShut
  {NoStop}%
%%CITATION = ARXIV:1408.3978;%%
\bibitem [{\citenamefont {Punturo}\ \emph {et~al.}(2010)\citenamefont {Punturo}
  \emph {et~al.}}]{et_punturo2010}%
  \BibitemOpen
  \bibfield  {author} {\bibinfo {author} {\bibfnamefont {M.}~\bibnamefont
  {Punturo}} \emph {et~al.},\ }\href
  {http://stacks.iop.org/0264-9381/27/i=8/a=084007} {\bibfield  {journal}
  {\bibinfo  {journal} {Classical and Quantum Gravity}\ }\textbf {\bibinfo
  {volume} {27}},\ \bibinfo {pages} {084007} (\bibinfo {year}
  {2010})}\BibitemShut {NoStop}%
\bibitem [{\citenamefont {Hild}\ \emph {et~al.}(2011)\citenamefont {Hild} \emph
  {et~al.}}]{Hild11}%
  \BibitemOpen
  \bibfield  {author} {\bibinfo {author} {\bibfnamefont {S.}~\bibnamefont
  {Hild}} \emph {et~al.},\ }\href
  {http://stacks.iop.org/0264-9381/28/i=9/a=094013} {\bibfield  {journal}
  {\bibinfo  {journal} {Classical and Quantum Gravity}\ }\textbf {\bibinfo
  {volume} {28}},\ \bibinfo {pages} {094013} (\bibinfo {year}
  {2011})}\BibitemShut {NoStop}%
\bibitem [{\citenamefont {{The ET Science Team}}(2011)}]{ET-D}%
  \BibitemOpen
  \bibfield  {author} {\bibinfo {author} {\bibnamefont {{The ET Science
  Team}}},\ }\href {http://www.et-gw.eu/index.php/etdsdocument} {\emph
  {\bibinfo {title} {Einstein gravitational wave Telescope conceptual
  design}}}\ (\bibinfo  {publisher} {European Commission},\ \bibinfo {year}
  {2011})\BibitemShut {NoStop}%
\bibitem [{\citenamefont {Grote}\ \emph {et~al.}(2013)\citenamefont {Grote},
  \citenamefont {Danzmann}, \citenamefont {Dooley}, \citenamefont {Schnabel},
  \citenamefont {Slutsky},\ and\ \citenamefont
  {Vahlbruch}}]{PhysRevLett.110.181101}%
  \BibitemOpen
  \bibfield  {author} {\bibinfo {author} {\bibfnamefont {H.}~\bibnamefont
  {Grote}}, \bibinfo {author} {\bibfnamefont {K.}~\bibnamefont {Danzmann}},
  \bibinfo {author} {\bibfnamefont {K.~L.}\ \bibnamefont {Dooley}}, \bibinfo
  {author} {\bibfnamefont {R.}~\bibnamefont {Schnabel}}, \bibinfo {author}
  {\bibfnamefont {J.}~\bibnamefont {Slutsky}},\ and\ \bibinfo {author}
  {\bibfnamefont {H.}~\bibnamefont {Vahlbruch}},\ }\href
  {https://doi.org/10.1103/PhysRevLett.110.181101} {\bibfield  {journal}
  {\bibinfo  {journal} {Phys. Rev. Lett.}\ }\textbf {\bibinfo {volume} {110}},\
  \bibinfo {pages} {181101} (\bibinfo {year} {2013})}\BibitemShut {NoStop}%
\bibitem [{\citenamefont {Tse}\ \emph {et~al.}(2019)\citenamefont {Tse} \emph
  {et~al.}}]{PhysRevLett.123.231107}%
  \BibitemOpen
  \bibfield  {author} {\bibinfo {author} {\bibfnamefont {M.}~\bibnamefont
  {Tse}} \emph {et~al.},\ }\href
  {https://doi.org/10.1103/PhysRevLett.123.231107} {\bibfield  {journal}
  {\bibinfo  {journal} {Phys. Rev. Lett.}\ }\textbf {\bibinfo {volume} {123}},\
  \bibinfo {pages} {231107} (\bibinfo {year} {2019})}\BibitemShut {NoStop}%
\bibitem [{\citenamefont {Acernese}\ \emph {et~al.}(2019)\citenamefont
  {Acernese} \emph {et~al.}}]{PhysRevLett.123.231108}%
  \BibitemOpen
  \bibfield  {author} {\bibinfo {author} {\bibfnamefont {F.}~\bibnamefont
  {Acernese}} \emph {et~al.},\ }\href
  {https://doi.org/10.1103/PhysRevLett.123.231108} {\bibfield  {journal}
  {\bibinfo  {journal} {Phys. Rev. Lett.}\ }\textbf {\bibinfo {volume} {123}},\
  \bibinfo {pages} {231108} (\bibinfo {year} {2019})}\BibitemShut {NoStop}%
\bibitem [{\citenamefont {Kimble}\ \emph {et~al.}(2001)\citenamefont {Kimble},
  \citenamefont {Levin}, \citenamefont {Matsko}, \citenamefont {Thorne},\ and\
  \citenamefont {Vyatchanin}}]{PhysRevD.65.022002}%
  \BibitemOpen
  \bibfield  {author} {\bibinfo {author} {\bibfnamefont {H.~J.}\ \bibnamefont
  {Kimble}}, \bibinfo {author} {\bibfnamefont {Y.}~\bibnamefont {Levin}},
  \bibinfo {author} {\bibfnamefont {A.~B.}\ \bibnamefont {Matsko}}, \bibinfo
  {author} {\bibfnamefont {K.~S.}\ \bibnamefont {Thorne}},\ and\ \bibinfo
  {author} {\bibfnamefont {S.~P.}\ \bibnamefont {Vyatchanin}},\ }\href
  {https://doi.org/10.1103/PhysRevD.65.022002} {\bibfield  {journal} {\bibinfo
  {journal} {Phys. Rev. D}\ }\textbf {\bibinfo {volume} {65}},\ \bibinfo
  {pages} {022002} (\bibinfo {year} {2001})}\BibitemShut {NoStop}%
\bibitem [{\citenamefont {Hild}\ \emph {et~al.}(2010)\citenamefont {Hild},
  \citenamefont {Chelkowski}, \citenamefont {Freise}, \citenamefont {Franc},
  \citenamefont {Morgado}, \citenamefont {Flaminio},\ and\ \citenamefont
  {DeSalvo}}]{Hild:2009ns}%
  \BibitemOpen
  \bibfield  {author} {\bibinfo {author} {\bibfnamefont {S.}~\bibnamefont
  {Hild}}, \bibinfo {author} {\bibfnamefont {S.}~\bibnamefont {Chelkowski}},
  \bibinfo {author} {\bibfnamefont {A.}~\bibnamefont {Freise}}, \bibinfo
  {author} {\bibfnamefont {J.}~\bibnamefont {Franc}}, \bibinfo {author}
  {\bibfnamefont {N.}~\bibnamefont {Morgado}}, \bibinfo {author} {\bibfnamefont
  {R.}~\bibnamefont {Flaminio}},\ and\ \bibinfo {author} {\bibfnamefont
  {R.}~\bibnamefont {DeSalvo}},\ }\href
  {https://doi.org/10.1088/0264-9381/27/1/015003} {\bibfield  {journal}
  {\bibinfo  {journal} {Class. Quant. Grav.}\ }\textbf {\bibinfo {volume}
  {27}},\ \bibinfo {pages} {015003} (\bibinfo {year} {2010})},\ \Eprint
  {https://arxiv.org/abs/0906.2655} {arXiv:0906.2655 [gr-qc]} \BibitemShut
  {NoStop}%
%%CITATION = ARXIV:0906.2655;%%
\bibitem [{\citenamefont {Mizuno}(1995)}]{Mizuno:1995iqa}%
  \BibitemOpen
  \bibfield  {author} {\bibinfo {author} {\bibfnamefont {J.}~\bibnamefont
  {Mizuno}},\ }\emph {\bibinfo {title} {{Comparison of optical configurations
  for laser-interferometric gravitational-wave detectors}}},\ \href@noop {}
  {Ph.D. thesis},\ \bibinfo  {school} {Hannover U.} (\bibinfo {year}
  {1995})\BibitemShut {NoStop}%
%%CITATION = INSPIRE-1228040;%%
\bibitem [{\citenamefont {Evans}\ \emph {et~al.}(2013)\citenamefont {Evans},
  \citenamefont {Barsotti}, \citenamefont {Kwee}, \citenamefont {Harms},\ and\
  \citenamefont {Miao}}]{Evans2013}%
  \BibitemOpen
  \bibfield  {author} {\bibinfo {author} {\bibfnamefont {M.}~\bibnamefont
  {Evans}}, \bibinfo {author} {\bibfnamefont {L.}~\bibnamefont {Barsotti}},
  \bibinfo {author} {\bibfnamefont {P.}~\bibnamefont {Kwee}}, \bibinfo {author}
  {\bibfnamefont {J.}~\bibnamefont {Harms}},\ and\ \bibinfo {author}
  {\bibfnamefont {H.}~\bibnamefont {Miao}},\ }\href
  {https://doi.org/10.1103/PhysRevD.88.022002} {\bibfield  {journal} {\bibinfo
  {journal} {Phys. Rev. D}\ }\textbf {\bibinfo {volume} {88}},\ \bibinfo
  {pages} {022002} (\bibinfo {year} {2013})}\BibitemShut {NoStop}%
\bibitem [{\citenamefont {Levin}(1998)}]{PhysRevD.57.659}%
  \BibitemOpen
  \bibfield  {author} {\bibinfo {author} {\bibfnamefont {Y.}~\bibnamefont
  {Levin}},\ }\href {https://doi.org/10.1103/PhysRevD.57.659} {\bibfield
  {journal} {\bibinfo  {journal} {Phys. Rev. D}\ }\textbf {\bibinfo {volume}
  {57}},\ \bibinfo {pages} {659} (\bibinfo {year} {1998})}\BibitemShut
  {NoStop}%
\bibitem [{\citenamefont {Freise}(2020)}]{Finesse}%
  \BibitemOpen
  \bibfield  {author} {\bibinfo {author} {\bibfnamefont {A.}~\bibnamefont
  {Freise}},\ }\href {http://www.gwoptics.org/finesse/} {\bibinfo {title}
  {\textsc{Finesse}: frequency domain interferometer simulation software}}
  (\bibinfo {year} {2020}),\ \bibinfo {note}
  {{\url{http://www.gwoptics.org/finesse}}}\BibitemShut {NoStop}%
\bibitem [{\citenamefont {Freise}\ \emph {et~al.}(2004)\citenamefont {Freise},
  \citenamefont {Heinzel}, \citenamefont {Lück}, \citenamefont {Schilling},
  \citenamefont {Willke},\ and\ \citenamefont {Danzmann}}]{Freise_2004}%
  \BibitemOpen
  \bibfield  {author} {\bibinfo {author} {\bibfnamefont {A.}~\bibnamefont
  {Freise}}, \bibinfo {author} {\bibfnamefont {G.}~\bibnamefont {Heinzel}},
  \bibinfo {author} {\bibfnamefont {H.}~\bibnamefont {Lück}}, \bibinfo
  {author} {\bibfnamefont {R.}~\bibnamefont {Schilling}}, \bibinfo {author}
  {\bibfnamefont {B.}~\bibnamefont {Willke}},\ and\ \bibinfo {author}
  {\bibfnamefont {K.}~\bibnamefont {Danzmann}},\ }\href
  {https://doi.org/10.1088/0264-9381/21/5/102} {\bibfield  {journal} {\bibinfo
  {journal} {Classical and Quantum Gravity}\ }\textbf {\bibinfo {volume}
  {21}},\ \bibinfo {pages} {S1067–S1074} (\bibinfo {year}
  {2004})}\BibitemShut {NoStop}%
\bibitem [{\citenamefont {Green}\ \emph {et~al.}(2017)\citenamefont {Green},
  \citenamefont {Brown}, \citenamefont {Dovale-{\'A}lvarez}, \citenamefont
  {Collins}, \citenamefont {Miao}, \citenamefont {Mow-Lowry},\ and\
  \citenamefont {Freise}}]{Green17}%
  \BibitemOpen
  \bibfield  {author} {\bibinfo {author} {\bibfnamefont {A.~C.}\ \bibnamefont
  {Green}}, \bibinfo {author} {\bibfnamefont {D.~D.}\ \bibnamefont {Brown}},
  \bibinfo {author} {\bibfnamefont {M.}~\bibnamefont {Dovale-{\'A}lvarez}},
  \bibinfo {author} {\bibfnamefont {C.}~\bibnamefont {Collins}}, \bibinfo
  {author} {\bibfnamefont {H.}~\bibnamefont {Miao}}, \bibinfo {author}
  {\bibfnamefont {C.~M.}\ \bibnamefont {Mow-Lowry}},\ and\ \bibinfo {author}
  {\bibfnamefont {A.}~\bibnamefont {Freise}},\ }\href
  {https://doi.org/10.1088/1361-6382/aa8af8} {\bibfield  {journal} {\bibinfo
  {journal} {Classical and Quantum Gravity}\ }\textbf {\bibinfo {volume}
  {34}},\ \bibinfo {pages} {205004} (\bibinfo {year} {2017})}\BibitemShut
  {NoStop}%
\bibitem [{\citenamefont {Buonanno}\ and\ \citenamefont
  {Chen}(2003)}]{Buonanno:2002mc}%
  \BibitemOpen
  \bibfield  {author} {\bibinfo {author} {\bibfnamefont {A.}~\bibnamefont
  {Buonanno}}\ and\ \bibinfo {author} {\bibfnamefont {Y.-b.}\ \bibnamefont
  {Chen}},\ }\href {https://doi.org/10.1103/PhysRevD.67.062002} {\bibfield
  {journal} {\bibinfo  {journal} {Phys. Rev.}\ }\textbf {\bibinfo {volume}
  {D67}},\ \bibinfo {pages} {062002} (\bibinfo {year} {2003})},\ \Eprint
  {https://arxiv.org/abs/gr-qc/0208048} {arXiv:gr-qc/0208048 [gr-qc]}
  \BibitemShut {NoStop}%
%%CITATION = GR-QC/0208048;%%
\bibitem [{\citenamefont {Th\"uring}\ \emph {et~al.}(2005)\citenamefont
  {Th\"uring}, \citenamefont {L\"uck},\ and\ \citenamefont
  {Danzmann}}]{PhysRevE.72.066615}%
  \BibitemOpen
  \bibfield  {author} {\bibinfo {author} {\bibfnamefont {A.}~\bibnamefont
  {Th\"uring}}, \bibinfo {author} {\bibfnamefont {H.}~\bibnamefont {L\"uck}},\
  and\ \bibinfo {author} {\bibfnamefont {K.}~\bibnamefont {Danzmann}},\ }\href
  {https://doi.org/10.1103/PhysRevE.72.066615} {\bibfield  {journal} {\bibinfo
  {journal} {Phys. Rev. E}\ }\textbf {\bibinfo {volume} {72}},\ \bibinfo
  {pages} {066615} (\bibinfo {year} {2005})}\BibitemShut {NoStop}%
\bibitem [{\citenamefont {Martynov}\ \emph {et~al.}(2019)\citenamefont
  {Martynov} \emph {et~al.}}]{Martynov:2019gvu}%
  \BibitemOpen
  \bibfield  {author} {\bibinfo {author} {\bibfnamefont {D.}~\bibnamefont
  {Martynov}} \emph {et~al.},\ }\href@noop {} {\bibfield  {journal} {\bibinfo
  {journal} {Phys. Rev.}\ } (\bibinfo {year} {2019})},\ \Eprint
  {https://arxiv.org/abs/1901.03885} {arXiv:1901.03885 [astro-ph.IM]}
  \BibitemShut {NoStop}%
%%CITATION = ARXIV:1901.03885;%%
\bibitem [{\citenamefont {Isogai}\ \emph {et~al.}(2013)\citenamefont {Isogai},
  \citenamefont {Miller}, \citenamefont {Kwee}, \citenamefont {Barsotti},\ and\
  \citenamefont {Evans}}]{Isogai:2013wfa}%
  \BibitemOpen
  \bibfield  {author} {\bibinfo {author} {\bibfnamefont {T.}~\bibnamefont
  {Isogai}}, \bibinfo {author} {\bibfnamefont {J.}~\bibnamefont {Miller}},
  \bibinfo {author} {\bibfnamefont {P.}~\bibnamefont {Kwee}}, \bibinfo {author}
  {\bibfnamefont {L.}~\bibnamefont {Barsotti}},\ and\ \bibinfo {author}
  {\bibfnamefont {M.}~\bibnamefont {Evans}},\ }\href
  {https://doi.org/10.1364/OE.21.030114} {\bibfield  {journal} {\bibinfo
  {journal} {Opt. Express}\ }\textbf {\bibinfo {volume} {21}},\ \bibinfo
  {pages} {30114} (\bibinfo {year} {2013})},\ \Eprint
  {https://arxiv.org/abs/1310.1820} {arXiv:1310.1820 [physics.optics]}
  \BibitemShut {NoStop}%
%%CITATION = ARXIV:1310.1820;%%
\bibitem [{\citenamefont {Kwee}\ \emph {et~al.}(2014)\citenamefont {Kwee},
  \citenamefont {Miller}, \citenamefont {Isogai}, \citenamefont {Barsotti},\
  and\ \citenamefont {Evans}}]{Kwee:2014vba}%
  \BibitemOpen
  \bibfield  {author} {\bibinfo {author} {\bibfnamefont {P.}~\bibnamefont
  {Kwee}}, \bibinfo {author} {\bibfnamefont {J.}~\bibnamefont {Miller}},
  \bibinfo {author} {\bibfnamefont {T.}~\bibnamefont {Isogai}}, \bibinfo
  {author} {\bibfnamefont {L.}~\bibnamefont {Barsotti}},\ and\ \bibinfo
  {author} {\bibfnamefont {M.}~\bibnamefont {Evans}},\ }\href
  {https://doi.org/10.1103/PhysRevD.90.062006} {\bibfield  {journal} {\bibinfo
  {journal} {Phys. Rev.}\ }\textbf {\bibinfo {volume} {D90}},\ \bibinfo {pages}
  {062006} (\bibinfo {year} {2014})},\ \Eprint
  {https://arxiv.org/abs/1704.03531} {arXiv:1704.03531 [physics.optics]}
  \BibitemShut {NoStop}%
%%CITATION = ARXIV:1704.03531;%%
\bibitem [{\citenamefont {Purdue}\ and\ \citenamefont
  {Chen}(2002)}]{Purdue:2002md}%
  \BibitemOpen
  \bibfield  {author} {\bibinfo {author} {\bibfnamefont {P.}~\bibnamefont
  {Purdue}}\ and\ \bibinfo {author} {\bibfnamefont {Y.}~\bibnamefont {Chen}},\
  }\href {https://doi.org/10.1103/PhysRevD.66.122004} {\bibfield  {journal}
  {\bibinfo  {journal} {Phys. Rev.}\ }\textbf {\bibinfo {volume} {D66}},\
  \bibinfo {pages} {122004} (\bibinfo {year} {2002})},\ \Eprint
  {https://arxiv.org/abs/gr-qc/0208049} {arXiv:gr-qc/0208049 [gr-qc]}
  \BibitemShut {NoStop}%
%%CITATION = GR-QC/0208049;%%
\bibitem [{\citenamefont {Degallaix}\ \emph {et~al.}(2019)\citenamefont
  {Degallaix}, \citenamefont {Michel}, \citenamefont {Sassolas}, \citenamefont
  {Allocca}, \citenamefont {Cagnoli}, \citenamefont {Balzarini}, \citenamefont
  {Dolique}, \citenamefont {Flaminio}, \citenamefont {Forest}, \citenamefont
  {Granata}, \citenamefont {Lagrange}, \citenamefont {Straniero}, \citenamefont
  {Teillon},\ and\ \citenamefont {Pinard}}]{degallaix_2019}%
  \BibitemOpen
  \bibfield  {author} {\bibinfo {author} {\bibfnamefont {J.}~\bibnamefont
  {Degallaix}}, \bibinfo {author} {\bibfnamefont {C.}~\bibnamefont {Michel}},
  \bibinfo {author} {\bibfnamefont {B.}~\bibnamefont {Sassolas}}, \bibinfo
  {author} {\bibfnamefont {A.}~\bibnamefont {Allocca}}, \bibinfo {author}
  {\bibfnamefont {G.}~\bibnamefont {Cagnoli}}, \bibinfo {author} {\bibfnamefont
  {L.}~\bibnamefont {Balzarini}}, \bibinfo {author} {\bibfnamefont
  {V.}~\bibnamefont {Dolique}}, \bibinfo {author} {\bibfnamefont
  {R.}~\bibnamefont {Flaminio}}, \bibinfo {author} {\bibfnamefont
  {D.}~\bibnamefont {Forest}}, \bibinfo {author} {\bibfnamefont
  {M.}~\bibnamefont {Granata}}, \bibinfo {author} {\bibfnamefont
  {B.}~\bibnamefont {Lagrange}}, \bibinfo {author} {\bibfnamefont
  {N.}~\bibnamefont {Straniero}}, \bibinfo {author} {\bibfnamefont
  {J.}~\bibnamefont {Teillon}},\ and\ \bibinfo {author} {\bibfnamefont
  {L.}~\bibnamefont {Pinard}},\ }\href
  {https://doi.org/10.1364/JOSAA.36.000C85} {\bibfield  {journal} {\bibinfo
  {journal} {J. Opt. Soc. Am. A}\ }\textbf {\bibinfo {volume} {36}},\ \bibinfo
  {pages} {C85} (\bibinfo {year} {2019})}\BibitemShut {NoStop}%
\end{thebibliography}
\end{document}